\documentclass[prb,superscriptaddress,twocolumn, showpacs]{revtex4-1}
\usepackage[normalem]{ulem}
\usepackage{cancel}
\usepackage{graphicx}
\usepackage{amssymb}
\usepackage{natbib}
\usepackage{dcolumn}
\usepackage{bm}
\usepackage{color}
\usepackage{amssymb,amsfonts,amsmath}
\usepackage{setspace}
\usepackage{mathrsfs}
\DeclareMathAlphabet{\mathpzc}{OT1}{pzc}{m}{it}
\def\bra#1{\mathinner{\langle{#1}|}} 
\def\ket#1{\mathinner{|{#1}\rangle}} 

\def\omw{\omega_{\mbox{\tiny MW}}}
\usepackage{comment}

\def\dip{U}

\newcommand{\HH}[1]{\hat{H}_{\text{#1}}}
\def\Hs{\HH{S}}
\def\Hb{\HH{R}}
\def\Hsb{\HH{S-R}}

\def\bsst{\beta_s^{\mbox{\tiny FM}}}
\def\hst{h^{\mbox{\tiny FM}}}

\newcommand{\titleinfo}{Thermalization and many-body localization in systems under dynamic nuclear polarization}

\newcommand{\Tr}{\operatorname{Tr}}

\def\pnans{p_n^{\mbox{\tiny Ans}}}
\def\pnstat{p_n^{\mbox{\tiny stat}}}
\def\pnstatp{p_{n'}^{\mbox{\tiny stat}}}

\graphicspath{{figures/}}

\begin{document}

\author{Andrea De Luca}
\affiliation{LPTMS, CNRS, Univ. Paris-Sud, Universit\'e Paris-Saclay, 91405 Orsay, France}
\author{In\'es Rodr\'iguez Arias}
\affiliation{LPTMS, CNRS, Univ. Paris-Sud, Universit\'e Paris-Saclay, 91405 Orsay, France}
\author{Markus M\"uller}
\affiliation{Condensed matter theory, Paul Scherrer Institute, CH-5232 Villigen PSI, Switzerland}
\affiliation{The Abdus Salam International Centre for Theoretical Physics, 34151, Trieste, Italy} 
\affiliation{Department of Physics, University of Basel, Klingelbergstrasse 82, CH-4056 Basel, Switzerland}
\author{Alberto Rosso}
\affiliation{LPTMS, CNRS, Univ. Paris-Sud, Universit\'e Paris-Saclay, 91405 Orsay, France}

\title{\titleinfo}
\begin{abstract}
We study the role of dipolar interactions in the standard protocol used to achieve dynamic nuclear polarization (DNP). 
In the so-called spin-temperature regime, where the interactions establish an effective 
thermodynamic behavior in the out-of-equilibrium stationary state, 
we provide numerical predictions for the level of hyperpolarization.
We show that nuclear spins equilibrate to the effective spin-temperature
established among the electron spins of radicals, 
as expected from 
the quantum theory of thermalization. 
Moreover, we present an analytical technique to estimate the spin temperature, and thus, the nuclear hyperpolarization in the steady state, as a function of interaction strength and quenched disorder. This reproduces
both our numerical data and experimental results. 
Our central finding is that the nuclear hyperpolarization increases steadily upon reducing the interaction strength (by diluting the radical density). Interestingly,
the highest polarization is reached at a point where 
the establishment of a spin temperature is just about to break down 
due to the incipient many-body localization transition in the electron spin system.  
\end{abstract}

\maketitle 

\section{General Introduction}
The phenomenon of many-body localization is currently attracting a lot of  attention, as it touches on various fundamental aspects of quantum statistical mechanics and quantum dynamics. However, despite its theoretical and conceptual appeal, very few  consequences of practical relevance for physical processes are known so far. In this paper,  we address a situation where  localization does play an important role. We analyze the effects of  incipient many-body localization in a system of driven quantum magnets, as standardly used  in preparing polarized nuclear spins. Interestingly, we find that the achieved nuclear polarization is optimized by tuning parameters very close to the localization transition, approaching it from the delocalized side, implying a practical aspect of the localization transition.

The canonical formulation of quantum statistical mechanics assumes the contact between the
system and an external reservoir. For a closed system, however, the 
description at large times using only few macroscopic parameters, 
such as the temperature or the chemical potential, implicitly assumes 
that the system itself can act as a thermal reservoir for its constituents.
Establishing the validity of this assumption and understanding the regimes where it breaks down constitute the still open problem 
of quantum thermalization \cite{Rigol2008, Polkovnikov2011}.
A simple way to probe thermalization is provided by quench protocols: a closed system is left to evolve
starting from a non-thermal
initial state, e.g. a state with a local excess of energy.
In infinite, ergodic systems the excess energy spreads and dilutes indefinitely, 
so that any memory of the initial imbalance is lost.  
If only conserved quantities are retained from the initial state, 
one can argue that two eigenstates which are globally similar (e.g. have the same energy),
cannot be distingushed by local measurements. 
This implies that expectation values and correlation functions of  local  observables in eigenstates 
must  coincide with their values in the microcanonical ensemble, 
a statement which  goes under the name of Eigenstate Thermalization Hypothesis 
(ETH)~\cite{deutsch1991quantum, *srednicki1994chaos}.

However, exceptions from such thermalizing systems can occur when the considered systems  
are sufficiently disordered,  
which may induce ergodicity breaking \cite{de2013ergodicity, deluca2014anderson} 
and the associated phenomenon of ``many-body localization'' (MBL) \cite{Anderson1958, Nandkishore2015}.
Recently, a variety of approaches based on perturbation theory \cite{Basko2006}, exact diagonalization \cite{Pal2010}, 
time-dependent DMRG \cite{Bardarson2012},
renormalization group \cite{Vosk2013a}, local integrals of motion~\cite{huse2014phenomenology, serbyn2013local, Ros2015} and even
rigorous mathematical results \cite{Imbrie2014}, provided independent indications of the existence of MBL phases.
In this localized phase, eigenstates have a very different structure with low entanglement 
following an area- rather than a volume-law \cite{bauer2013area}. In particular they
do not obey the ETH, reflecting that the quantum dynamics is not ergodic anymore:  local 
expectation values exhibit strong fluctuations from eigenstate to eigenstate of the same energy density,
since an extensive set of parameters \cite{chandran2015constructing, Ros2015} is necessary to describe the long-time dynamics. 

So far only few experimental indications of many body localization have been reported, in cold
atoms \cite{schreiber2015observation} and trapped ions \cite{smith2015many}, where the difficulty 
of isolating quantum systems from a thermalizing bath can be overcome 
more easily than in solid matter which always hosts phonons.
However, the possibility of hole burning in frustrated magnets such as LiHo$_x$Y$_{1-x}$F$_4$ and Gadolinium Gallium Garnet~\cite{ghosh2002coherent} 
suggests that quantum magnets are promising solid state systems where localization 
phenomena might manifest themselves over very long time scales. 
After all, it was the apparent absence of spin diffusion in disordered magnets~\cite{feher1959electron} 
that had led Anderson~\cite{Anderson1958} to start the investigation of localization physics over half a century ago.

Being in a localized phase is often considered interesting for quantum technological applications, 
since entanglement is limited, or grows only very slowly in time~\cite{serbyn2013universal, bauer2013area, chandran2014many}. 
In this work, we show that being close to a localization transition 
is  of great interest also in an entirely different domain, as it helps to achieve large nuclear hyperpolarization in quantum magnets.

In this article, we study a class of quantum magnets relevant for so-called dynamic nuclear polarization (DNP). 
Those magnets generically host electron spins in a more or less random spatial configuration. 
To achieve a hyper-polarization the system is driven externally by microwaves.
Apart from their immediate interest for DNP, these systems constitute an interesting example where the approach of a localized regime impacts the {\em steady state} of 
driven systems -- rather than studying eigenstates of time-independent problems. 
Note that in general, even in an ergodic regime with  weak disorder, no simple thermodynamic description of a driven
steady state can be  expected. However, in the presence of weak driving and coupling with the outside world, the  dephasing 
time of internal degrees of freedom is much faster than any other time-scale, including the one associated with the driving. Hence  
the density-matrix becomes essentially diagonal in the basis of the many-body eigenstates before the effect of the driving 
or the environment is felt. 
In this limit, if ETH holds for the isolated system, the driven state can nevertheless be described to a good approximation by the same set of intensive parameters as in equilibrium, 
however, with values that depend on the driving. In contrast, once localization occurs in the closed spin system, 
the steady state reflects the details of the local dynamics of the drive and 
a simple equilibrium-like characterization will not emerge in general. 
The failure of an effective equilibrium description could potentially be investigated 
as a fingerprint of localization physics.

In this work we present clear signatures of thermalization and many-body localization
in a model describing the hyperpolarization of nuclear spins obtained via 
dynamic nuclear polarization (DNP) \cite{Abragam1982a}. 
\begin{figure}[t]
\includegraphics[width=\columnwidth]{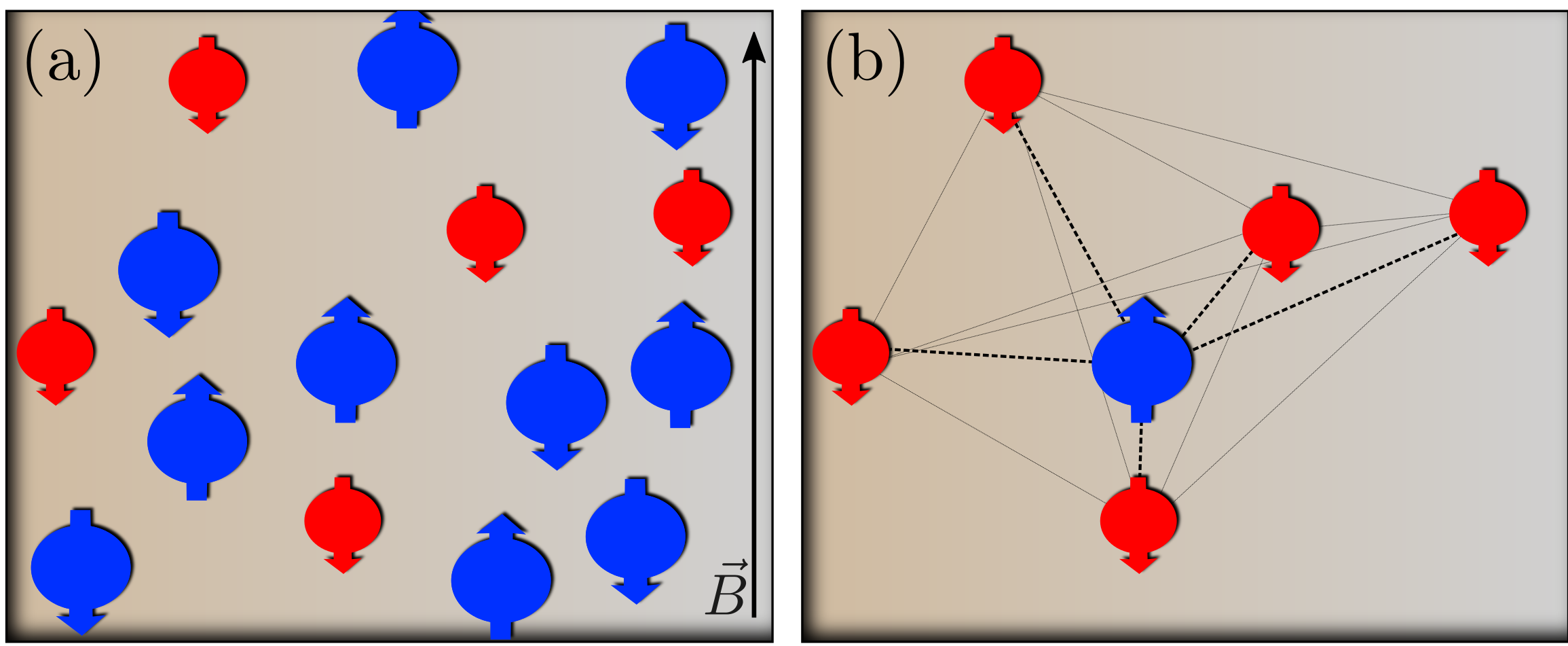}
\caption{\label{DNPsketch}
Color online. Left: DNP system: the spatial positions of the nuclear spins 
of the compound (in blue) and the electron spins of few radical molecules (in red)  are frozen 
in glassy matrix. The spins are coupled via dipolar and hyperfine interactions. 
Right: The simplified model of  Eq.~\eqref{hamiltonian}. A single nuclear spin 
is surrounded by a collection of electron spins. Each spin is assumed to have random interactions with all others. 
The electro-nuclear couplings are much weaker than the dipolar couplings connecting the electron spins among each other.
}
\end{figure}
For a typical DNP procedure, one works with a compound doped with radicals 
(i.e., molecules with unpaired electrons), which is rapidly quenched to low temperatures to form a frozen, glassy matrix.
\footnote{Empirically a glassy atomic structure is necessary to obtain significant hyper-polarization. 
The reason is not entirely established, but such a structure presumably helps to ensure a homogeneous 
dilution of radicals (and to prevent the clustering of radicals at lower-dimensional grain boundaries).}
Such a compound is then exposed to a strong magnetic field $|\vec B| \simeq 3$ T, put in contact 
with a cold reservoir at $\beta^{-1} \simeq 1 $K (see Fig.~\ref{DNPsketch} Left), and finally irradiated with microwaves.
In the absence of microwave irradiation, the system reaches thermal equilibrium at the reservoir temperature
$\beta^{-1}$: the unpaired electrons are strongly polarized ($\sim 94\%$) while the nuclear spins are very weakly
polarized ($< 1 \%$) as the nuclear Zeeman gap is about three orders of magnitude smaller than the level splitting of the electrons.
In contrast, when the microwave frequency is close to the electron's Zeeman gap, 
the driven system of interacting electron and nuclear spins organizes
into an out-of-equilibrium steady-state with a huge nuclear polarization. 
The hyperpolarized sample  can  then  be  dissolved  at  room temperature 
\cite{ardenkjaer2003increase}, injected  in  patients,  and  used  as  metabolic  tracer \cite{Golman2006}.

The traditional explanation of hyperpolarization goes back to the seventies and is based on the idea that the spin system, when irradiated, cools down to an effective thermodynamic state characterized by a low spin-temperature~\cite{Provotorov1962, Abragam1982a}.
This idea has been qualitatively confirmed in several experiments  \cite{lumata2011, kurdzesau2008}, 
where the enhanced polarizations of different nuclear species ($^{13}C$, $^{15}N$, $^{89}Y$, $\ldots$)  
are  well described by an equilibrium-like formula,  
\begin{equation}
\label{TM}
P_n = \tanh(\beta_s \hbar \omega_n /2),
\end{equation}
  with a unique parameter $\beta_s$ (the inverse spin temperature), but different Zeeman gaps $\omega_n$ for the different species.

The success of the spin-temperature picture raises two fundamental questions:
\begin{itemize}
 \item How can the emergence of an effective thermodynamical description 
  be justified on  a microscopic basis, and under what circumstances does an effective equilibrium indeed occur in the steady state?
 \item How to estimate and optimize $\beta_s$ as a function
 of the microscopic parameters which can be controlled in an experiment (e.g., radical concentration, magnetic field strength,
 microwave intensity, \ldots)?
\end{itemize}
As for the first question, we will see that an effective equilibrium does not emerge in just any spin system. Rather, 
the thermalization tendency in the closed spin system has to be sufficiently strong in order for an effective spin temperature to establish. This in turn poses constraints on the relative strength of disorder and interactions,
which can be tuned independently in experiments, as we will discuss.
Previous attempts to predict the value of $\beta_s$ were based on a purely phenomenological approach where 
an out-of-equilibrium quasi-thermal state was postulated from the outset, without questioning in which parameter regime the hypothesis was actually justified.
In typical experimental settings, the spin temperature was estimated to be three orders of magnitude smaller than the lattice temperature. However, the theory predicted 
largely unrealistic levels of hyperpolarization:
up to $80\%$, as compared to the $20-40\%$ observed in actual experiments \cite{borghini1968spin}.

In a recent work, we studied an ensemble of electron spins with
random Zeeman gaps and subject to dipolar interactions, but with no nuclear spins \cite{DeLuca2015}. 
Under microwave irradiation and assuming a weak coupling 
to the reservoir, we showed that the quantum dynamics can be reduced
to a master equation 
for the occupation probabilities 
of the interacting eigenstates 
\cite{hovav2010theoretical, *Hovav2012, *Hovav2013}. For $N=12$ spins, the (unique) stationary state can be extracted numerically  
from the master equation: the steady polarizations of the electrons reflect the  ergodicity properties of eigenstates. 
In particular we have shown that an effective spin-temperature emerges whenever: i) ETH is satisfied by
the isolated spin system; ii) dephasing inside the system happens on a fast time-scale as compared to the 
driving and the relaxation processes 
involving the bath.
In contrast, the signatures of a spin temperature disappear when the electron spins are many-body localized. 

In this work, we study numerically the hyperpolarization of a single nuclear spin 
interacting with all the electron spins (see Fig.~\ref{DNPsketch} Right). 
We show that its stationary polarization is consistent with the 
spin-temperature prediction of Eq.~(\ref{TM}) when the spin Hamiltonian obeys ETH. 
This indicates that the steady-state 
parameter $\beta_s^{-1}$
behaves as a genuine temperature. 
Then, in the framework of the master equation already introduced in [\onlinecite{DeLuca2015}], 
we derive a technique to estimate $\beta_s$. 
The microscopic details of the system Hamiltonian
are  encoded only in the equilibrium spin-spin correlation function,
which can  be computed analytically. The results are in good agreement
both with our numerical data and, qualitatively, with experimental measurements. 
This paves the way to finding a set-up to optimize the hyperpolarization efficiency.
We find that by decreasing the radical concentration, the spin-temperature is lowered 
(and hence the hyperpolarization is increased)
until the stationary nuclear polarization reaches a maximal value. 
Below this threshold concentration
an approach postulating thermalization and an effective spin-temperature would still predict 
a monotonous continuation of these trends. However, at this point
the electron spins actually start to localize, thermalization is impeded and as a consequence 
the average hyperpolarization becomes strongly suppressed.
The maximum efficiency in hyperpolarization is thus a fingerprint of the incipient MBL transition, 
which occurs as the radical concentration is reduced or the external magnetic field is increased. 
Since both parameters can be controlled in standard DNP experiments, these theoretical predictions can be directly 
subjected to experimental verification.

The paper is organized as follows: in Sec.~\ref{secDNPintro} we introduce the DNP protocol and 
 detail how we model it. In particular we discuss how  the nuclear spin is coupled to the electron spins. In Sec.~\ref{secSpinTemp}, we show how the  spin temperature
can be defined for the stationary state and how its value can be computed numerically and analytically; in Sec.~\ref{secResults}
we present our results while  Sec.~\ref{secdiscussion} discusses the applicability and the breakdown of the spin-temperature Ansatz 
as the closed system undergoes an MBL transition.

\section{The DNP setting and a simplified model \label{secDNPintro}}

The enhancement of the nuclear polarization emerges in the framework of a correlated quantum spin system
driven far from equilibrium by resonant microwave irradiation, while being in contact with the thermal reservoir 
of atomic lattice degrees of freedom, held at a temperature of typically about $T=1$K.
Describing the dynamical behavior of such a system is a formidable task, due to  several competing interactions. Below, we specify the various ingredients and 
derive a simplified model that  can be investigated both numerically and analytically.

\subsection{Description of the system}
In a DNP set-up, there are three fundamental ingredients: i)
the internal spin dynamics governed by the competition between disorder and interactions; ii)
the microwave pumping; iii) the weak contact with the reservoir.
Let us describe each of them in turn.

\textit{The spin Hamiltonian. ---} 
The spin system is composed of the electron spins ($S_i$, $i=1,..., N$) associated with the radical molecules, and nuclear spins ($I_j$, labelled $j=1,...,N_n$)  that one aims to
hyperpolarize.
Typically the  concentration of the nuclei is about $N_n/N \approx 10^3$ times larger than that of the radicals. For simplicity we will consider both electron and nuclear spins to have spin $1/2$.  
All spins are exposed to a strong uniform magnetic field.  
The system Hamiltonian then takes the form
\begin{eqnarray}
 \hat H_{\mbox{\tiny S}} &=& \hat H_Z+\hat H_{\mbox{\tiny int}},\nonumber\\
 \hat H_Z &=&  \sum_{i =1}^N \left( \omega_e +\Delta_i \right) \hat S_z^i - \omega_n \sum_{j=1}^{N_n} \hat I_z^j,
 \label{zeeman}
\end{eqnarray}
where $\omega_e, \omega_n$ describe the average Zeeman energy of electron and nuclear spins in the external magnetic field, respectively. Because of  $g$-factor anisotropies \cite{Abragam1982a}, the electronic Zeeman energies 
are subject to spatially fluctuating contributions $\Delta_i$, which constitute the dominant source of quenched disorder in the problem.

The interaction term $\hat H_{\mbox{\tiny int}}$
includes three contributions
\begin{itemize}
 \item the dipolar interaction between electron spins, of the form
 \begin{equation}
  \label{fulldipolar}
  \hat H_{e-e} = \sum_{i<j} \frac{\mu_0 \gamma_e^2}{4\pi |\mathbf{r}_{ij}|^3} \left[\hat{\mathbf{S}}^i \cdot \hat{\mathbf{S}}^j - 
   3( \hat{\mathbf{S}}^i \cdot \mathbf{n}_{ij})( \hat{\mathbf{S}}^j \cdot \mathbf{n}_{ij})\right],
 \end{equation}
where $\mathbf{r}_{ij}$ is the distance vector between spins $i$ and $j$, $\mathbf{n}_{ij} = \mathbf{r}_{ij}/|\mathbf{r}_{ij}|$
and $\Gamma_e $ is the coupling constant. In a large magnetic field, this term
can be regarded as a perturbation of the Zeeman energy described by \eqref{zeeman}. Therefore, hybridization
between sectors of different total electronic magnetization $\hat S_z = \sum_{i} \hat S_z^i$, is strongly suppressed by the 
Zeeman gap. With a typical distance between radical molecules of $r_{e-e} \simeq 20 \AA$,
the smallness of the parameter 
$\mu_0 \gamma_e^2/r_{e-e}^3 \omega_e \simeq 10^{-3}$
justifies the secular approximation  \cite{Abragam1982a}, which projects the Hamiltonian onto these subspaces,
\begin{equation}
  \label{secdipolar}
  \hat H_{e-e} = \sum_{i<j} \dip_{ij} \left[4 \hat{S}^i_z \hat{S}^j_z -(\hat{S}^i_+ \hat{S}^j_- + 
  \hat{S}^i_- \hat{S}^j_+)\right],
 \end{equation}
 with $\dip_{ij} = \mu_0 \gamma_e^2(1 - 3 \cos^2\theta_{ij}) /(16\pi|\mathbf{r}_{ij}|^{3})$. 
 Here, $\theta_{ij}$ is the angle between the field along $z$ and $\mathbf{r}_{ij}$, 
 and $[\hat S_{+}^i, \hat S_{-}^i] = 2 \hat S_z^i$, with $\hat S^i_\pm = \hat S^i_x \pm i \hat S^i_y$.
 \item the dipolar interaction between nuclear spins, which takes the same form as in \eqref{fulldipolar} with $\gamma_e \to \gamma_n \simeq 10^{-3} \gamma_e$. These interactions are responsible for nuclear spin-diffusion\cite{karabanov2015dynamic}, 
 which tends to homogenize the polarization among  the nuclear spins. 
 \item the hyperfine interaction between the nuclear and the electron spins. 
 For large Zeeman fields a projection onto $S^z$ preserving terms yields
  \begin{equation}
  \label{sechf}
  \hat H_{e-n} = \sum_{i,j} D_{ij}^{(z)} 
  \hat{S}_z^i \hat{I}_z^j + D^{(x)}_{ij} 
  \hat{S}_z^i \hat{I}_x^j  +D^{(y)}_{ij} \hat{S}_z^i \hat{I}_y^j.
 \end{equation}
Note that one has the freedom to rotate the nuclear spin along the $z$-axis and therefore by 
a unitary transformation, we can always set $D^{(y)}=0$. The hyperfine couplings are dominated by
their anisotropic component and $D^{(x)}_{ij} \simeq \mu_0 \gamma_e \gamma_n |r_{e-n}^{-3}|$, with $r_{e-n}$
the distance between electrons and nuclei. Again, the perturbative treatment 
is justified by the small value of the hyperfine coupling
$D^{(x)}_{ij}/ \omega_e \simeq 10^{-5} \div 10^{-8}$. 
\end{itemize}

\textit{Coupling to the lattice. ---}
The bath modes must be included in the Hamiltonian $\hat{\mathcal{H}}$
of the full set-up by adding a priori two terms
\begin{equation}
\label{latticeHam}
 \hat{\mathcal{H}} = \Hs + \Hb + \Hsb.
\end{equation}
Here $\Hb$ describes the dynamics of the bath, i.e., the motion of the atoms, and
$\Hsb$ captures the spin-bath interaction.
The simplest description of such an interaction is obtained by assuming that each spin couples  to the displacement  of  one individual
vibrational mode localized close to it,
\begin{equation}
\label{spinbath}
\Hsb =  \lambda_e \sum_{\stackrel{i=1}{\alpha = x,y,z}}^N \hat S^i_\alpha \hat \Phi^i_{\alpha,e}
+\lambda_n \sum_{\stackrel{j=1}{\alpha = x,y,z}}^N \hat I^j_\alpha \hat \Phi^j_{\alpha,n}\;.
 \end{equation}
Here $\lambda_e, \lambda_n$ are the electron and nuclear spin-bath coupling constants, respectively. They fix the time-scale of the
relaxation processes. Since $\lambda_n \ll \lambda_e$, nuclear spin-bath coupling does
not play any role in the hyperpolarization procedure.
As usual the specific details of the bath Hamiltonian $\Hb$ are unimportant, its main role being  to maintain the bath at temperature $\beta^{-1} \equiv T$ and to quickly erase the memory of its past interactions with the system.

\textit{The microwave pumping. ---}
The microwave frequency $\omw$ is tuned close to the average electronic Zeeman energy $\omega_e$, so that it manages to flip electron spins occasionally, one at a time. This is described by the time-dependent Hamiltonian
\begin{equation}
 \label{hammw}
\hat H_{\mbox{\tiny MW}}(t) = 2 \omega_1 \sum_i \hat S_x^i \cos(\omw t),
\end{equation}
where $\omega_1$ is the amplitude
of the microwave field.

One of the central experimental observables is the so-called DNP profile, which describes the stationary value of nuclear polarization as a function of $\omw$ (see Fig.~\ref{dnpprofile}).
In order to obtain a sizeable enhancement over the thermal nuclear spin polarization,
the microwave frequency $\omw$ must be chosen such as to lie within the range of the inhomogeneously broadened spectrum of Zeeman energies, $(\omw - \omega_e)^2 \lesssim \overline{\Delta_i^2}$,
so as to be resonant with a fraction of electron spin-flip transitions.

\subsection{Simplified model}

In this article, we study a model of $N_n=1$ nuclear spin and $N$ interacting electron spins, 
in contact with a thermal reservoir and driven out-of-equilibrium
by microwave irradiation. 
We can limit the study to a single nuclear spin,
because of an experimental fact observed under standard DNP conditions.
A sample of $^{13}C$ pyruvic acid was doped with trytil radicals, a stable and very efficient polarizing agent.
The choice of trytil is such that $\overline{\Delta_i^2} \ll \omega_n^H$, so that hydrogen is not DNP active. 
In this way, the only active nuclear species is $^{13}C$ and one observes that changing the nuclear spin concentration of $^{13}C$ with respect to the spinless $^{12}C$,
does not affect the final polarization of $^{13}C$ as shown in [\onlinecite{ColomboSerra2014}]. 
This suggests that the enhancement in the nuclear polarizations is inherited from the steady state of the electrons
and that different nuclei always have a homogeneous polarization: 
therefore, the study of a single nuclear spin should suffice to demonstrate the 
existence of a spin temperature and its transfer to the nuclear spins.

DNP is empirically found to be effective only in compounds
where the atoms are frozen into a random, glassy configuration in which
the distances $\mathbf{r}_{ij}$ between pairs of radicals
are random. Since we will study relatively small systems, we model this in  
the limit of fully connected spin-spin interactions taking $\dip_{ij}$ as Gaussian random variables 
with zero mean and variance $U^2/N$. Note that the simplest choice of non-fluctuating couplings $U_{ij} = U/N$
would be pathological as it leads to the integrable Richardson model, which is always non-thermal\cite{Buccheri2011}.

Further, we neglect the Ising coupling $\hat{S}^i_z \hat{S}^j_z$ in \eqref{secdipolar}, 
as it mostly modifies the instantaneous local fields seen by the various electron spins, 
but does not contribute essentially to the physics.

The inhomogeneous contributions to the Zeeman energy, $\Delta_i$,   
are taken 
equally spaced inside the interval $[-\Delta\omega_e, \Delta\omega_e]$ \footnote{Explicitly, the energy shifts are chosen as
$\Delta_i = \Delta \omega_e \left( \frac{2i -  N-1}{N} \right)$ with $i = 1,\ldots, N$.}. 
A more realistic approximation would consider a random distribution of the $\Delta_i$. 
However this choice would lead to strong finite size fluctuations, since in the small systems accessible
by numerics, at most a single electron is in resonance with the microwave irradiation.

We finally arrive at the following simplified Hamiltonian:
\begin{multline}
 \label{hamiltonian}
 \hat H_{S} =  \sum_{i =1}^N \left( \omega_e +\Delta_i \right) \hat S_z^i 
+ \sum_{i<j} U_{ij} (\hat{S}^i_+ \hat{S}^j_- + 
  \hat{S}^i_- \hat{S}^j_+) + \\ 
  - \omega_n \hat I_z + \sum_i D_{i} \hat S^i_z  \hat I_x.
 \end{multline}
Here we retain only one representative term for the electron-nuclear spin coupling, which induces flips of the nuclear spin and leads to its quasi-thermalization, if the electron spins establish a spin temperature.
The $D_i$'s are drawn from a 
normal distribution of zero average and variance $D^2/N$. The strength $D$ 
is chosen so as to ensure sufficient coupling to the electrons without causing
any significant perturbation of their state. The nuclear spin thus acts effectively as a thermometer. 

Let us start by considering the relaxation dynamics in the absence of microwaves. 
In order to treat the interaction with the reservoir, we employ a significant separation of 
time-scales in our problem. As we observed in the previous section, 
the leading term in the Hamiltonian is proportional to $\omega_e \simeq 100$ GHz in typical magnetic fields. The order of magnitude of the remaining terms of the spin Hamiltonian 
$(\Delta \omega_e, U, b, \omega_n)$ range from a few to $100$ MHz. 
On the other hand, the rate of energy exchange with the bath can be estimated by the inverse 
of the electronic relaxation time, $1/T_{1e}$,
which is of the order of $1$ Hz~\cite{filibian2014role}: this implies a \textit{weak coupling} 
between electron spins and bath modes. 
In this limit, it is possible to derive, within the Born-Markov approximation scheme,
an evolution equation for the density-matrix in Lindblad form. 
This approach is
analogous to the equations derived by [\onlinecite{hovav2010theoretical}] and has the advantage of being 
directly connected to the microscopic model \cite{petruccione2002theory}.
The details of this derivation were presented in [\onlinecite{DeLuca2015}].
Summarizing, the resulting evolution of the density matrix of the spin system, $\rho$, has the Lindblad form
\begin{equation}
\label{lindblad}
 \frac{d \rho}{dt} = -i [\hat H_S, \rho] + \mathcal{L}[\rho]\;.
\end{equation}
The last term contains the non-unitary dynamics with two kinds of contributions: i) electron spin-flip processes (due to the coupling of bath modes with 
$S_x^i, S_y^i$ in \eqref{spinbath}) which induce transitions between pairs of eigenstates of $\hat H_{S}$
and involve an exchange of energy $\simeq \omega_e$ with the lattice; 
ii) processes due to the coupling of the bath modes with $\hat S_z^i$ or $\hat I_z$ 
for which the exchange of energy vanishes in the non-interacting limit.
The latter thus contributes mostly to the dephasing of off-diagonal elements
of the density-matrix in the basis of eigenstates of $\hat H_S$, 
while the former dominates the relaxation of the diagonal elements, so that 
at long-times the density-matrix reaches the thermal state at the lattice temperature: 
$\rho = Z^{-1} e^{-\beta \hat H_S}$.

The precise estimation of the different time scales is difficult. 
Experimentally, one knows that $T_{1e} \simeq 1$ s 
and $T_{2e} \simeq 10^{-6}$ s, associated with the relaxation of the longitudinal 
and the transverse spin polarization, respectively \cite{Johannesson2009}.
Given the huge difference between the two time scales \footnote{Indeed, the electron spin-flips induced by the bath are the slowest process and fully determine the polarization time, $T_{pol}=\frac{N_n}{N_e}T_{1e}$\cite{Filibian2014}} we assumed in Ref. [\onlinecite{DeLuca2015}] that 
the quantum dynamics of the system can be reduced to 
a classical master equation for the occupation probabilities $p_n$ of the many-body eigenstates $\Psi_n$ 
\begin{equation}
 \label{mastereq1}
\frac{dp_n}{dt} = \sum_{n' \neq n}  W_{n' \to n} p_{n'} - W_{n\to n'} p_n \;,
\end{equation}
where the transition rate has the form $W_{n,n'} = W^{\text{bath}}_{n, n'}$:
\begin{equation}
\label{matrixtransitionBATH}
W^{\text{bath}}_{n, n'} =  \frac{2h_\beta(\Delta \epsilon_{n,n'}) }{T_{1e}}\sum_{j=1}^N \sum_{\alpha=x,y,z} |\bra{n}\hat S_{\alpha}^{j}\ket{n'}|^2\;,\\
\end{equation}
Eq.~\eqref{matrixtransitionBATH} describes spin-flips induced by the interaction with the external 
bath on a time scale $T_{1e}$ and
the function $h_\beta(x)=e^{\beta x}/(1+ e^{\beta x})$ 
assures detailed balance and convergence to Gibbs equilibrium at temperature $\beta^{-1}\equiv T \simeq 1$K.
The coupling of the nuclear spin to the bath modes 
induces similar transitions but with a much smaller rate as 
$T_{1n} \gtrsim 10^3 T_{1e}$.
We remark that a single electron spin would dephase also in absence of 
the external reservoir due to dipolar coupling with the other spins. 
In particular, in the ergodic phase, the system can act as its own reservoir 
and there is an internal notion of dephasing: the experimental $T_{2e}$ is affected by the internal one, but for
simplicity treat it as a separate input parameter independent of the interaction strength $U$.

We now include the coupling to the microwave radiation as described by \eqref{hammw}.
As the microwave field is time-dependent,
it requires some care. A precise estimation of the microwave amplitude 
$\omega_1$
is difficult, since it is hard to evaluate the fraction of emitted
power which actually reaches the sample in a given experiment. In general, we can assume that 
$\omega_1$ is within the range of tens to hundreds of kHz. 
Thus, the transition rate satisfies $\omega_1^2 T_{2e}\ll \omega_e$. Together with the condition  $\Delta \omega_e \ll \omega_e$ this ensures that we can safely employ the well-known \textit{rotating-wave approximation}: 
It entails transforming the Hamiltonian and the density-matrix into a rotating frame, i.e., 
\begin{align}
 \rho^{(r)} &= e^{i \hat S_z \omw t} \,\rho\, e^{-i \hat S_z \omw t} \;, \label{rhorot}\\
 \Tr[ \hat O \rho] &= \Tr[ e^{- i \hat S_z \omw t} \hat O e^{i \hat S_z \omw t} \rho^{(r)}],
\end{align}
where the last line holds for any observable $\hat O$, and $\hat S_\alpha = \sum_i \hat S^i_\alpha$. For observables that commute with $\hat S_z$, such as the individual polarizations $\hat S_z ^i$, expectation values can safely be computed in the rotating frame.
The advantage of the transformation \eqref{rhorot} is that, since $[\hat S_z, \hat H_S ] = 0$, the evolution of $\rho^{(r)}$,
in the presence of microwaves, is the same
as that for $\rho$ in \eqref{lindblad}, 
where, apart from rapidly oscillating terms, the Hamiltonian has been replaced by the time-independent
\begin{equation}
\hat H_S^{\mbox{\tiny rot}} \to \hat H_S - \omw \hat S_z + \omega_1 \hat S_x 
\end{equation}
while $\mathcal{L}[\rho] \to \mathcal{L}[\rho^{(r)}] $.
Within this approximation the effect of microwaves can be included 
in the master equation \eqref{mastereq1} as an additional rate
$W_{n,n'} = W^{\text{bath}}_{n, n'} + W^{\text{MW}}_{n, n'}$, with
\begin{equation}
\label{matrixtransitionMW}
W^{\text{MW}}_{n, n'} = \frac{4 \omega_1^2 T_{2e} |\bra{n}  \hat S_x \ket{n'}|^2}{1+ T_{2e}^2 (|\epsilon_n - \epsilon_{n'}| - \omw )^2}\;.
\end{equation}
Again, the use of the master equation is justified as long as $\omega_1^2 T_{2e}$ 
is small as compared to the amplitude of the intra-spin interaction terms.

It is important to observe that for $\omega_1 \neq 0$,
the rates $W_{n,n'}$ do not respect a detailed balance condition and thus, 
the  stationary state will be out-of-equilibrium.

\section{The spin-temperature Ansatz for the stationary state   \label{secSpinTemp}}

In this section we discuss the behavior of the system ignoring the nuclear spin, which is weakly coupled and serves 
only as a thermometer, without acting back on the electronic system.   
The time evolution of the spin system can be decomposed into two regimes. At rare times, the thermal reservoir or the microwave field flips a single electron spin. 
Subsequently,  
fast dephasing brings the system essentially into a classical 
mixture of eigenstates of $\hat H$ (this is a good description as far as  local observables are concerned).
Since the spin-flip is a perturbation localized in space,
it is natural to ask how much information about the position of the flipped spin is retained 
after dephasing. As long as the eletron spins form an ergodic system no local information 
except for the increment of the conserved quantities (energy and electron spin polarization)
will remain. It implies that after the typical dephasing time, for any local observable, the expectation value on the projected state coincides 
with the average over all states characterized by the same value of energy and electron polarization. 
In a canonical description this corresponds to the equilibrium average in presence of two 
intensive parameters: 
\begin{equation}
\label{spintempAnsatz}
p_n \simeq \pnans = Z^{-1} e^{-\beta_s(\epsilon_n + h s_{z,n})} \;,
\end{equation}
where $Z$ is fixed by normalization. 
The inverse spin-temperature, $\beta_s$, is the parameter conjugate to the  energy, 
while the effective magnetic field, $h$, is conjugate to the electron magnetization:
$\epsilon_n$ and  $s_{z,n}$ are the eigenvalues of $\hat H$ and $\hat S_z$ on $\ket{n}$ in the laboratory (non-rotating) frame. 
In the following we compute both parameters following two complementary approaches: 
\begin{itemize}
\item A fitting method (FM), which allows us to infer $\beta_s$ and $h$ 
from numerical simulations, by imposing that the distribution
$\pnans$ in \eqref{spintempAnsatz} has the exact average energy and electron magnetization.
\item A perturbative expansion (PE) 
 for weak $U$ and for an infinite number of spins interacting via (\ref{hamiltonian}). 
This method is based on the observation that the variation of energy and electron polarization, 
induced by single spin flip transitions, are encoded in
the spin-spin correlation functions computed with the density-matrix at time $t$.
Using (\ref{spintempAnsatz}), these functions can be computed, at least order by order in perturbation theory in $U$.
The values of $\beta_s$ and $h$ can then be determined by imposing a balance for the total in- and outflow of energy and electron spin polarization due to interactions with the radiation and the bath.
Note that the thermal average of correlation functions is only weakly affected by localization at small $U$ and therefore
perturbation theory can produce sensible results. 
\end{itemize}
We now explain the details of the two procedures.

\subsection{Fitting method \label{spintempSEC}}
Our fitting method allows us to fix the parameters of the spin-temperature Ansatz
for a given value of $U$.
We use a numerical simulation to obtain the stationary state 
of Eq.~\eqref{mastereq1} in a given realization of the couplings $\dip_{ij}$ and $D_i$. 
First, by exact diagonalization of the Hamiltonian $\hat H$, 
we compute the $2^{N+1}$ eigenstates $\ket{n}$ of energy $\epsilon_n$ and total electron polarization $s_{z,n}$. 
The exponential growth of the Hilbert space strongly limits the accessible sizes, and thus we restrict ourselves to 
$N=12$. The rates in Eqs.~(\ref{matrixtransitionBATH},~\ref{matrixtransitionMW}) can be computed exactly 
as matrix elements between pairs of eigenstates. 
Then, the occupation probabilities in the stationary state $\pnstat$ are obtained by
setting $dp_n/dt = 0$ in Eq.~(\ref{mastereq1}) and solving the resulting linear system. Typically there is a unique solution to these equations, both in the ergodic and the localized phases of the isolated system. In particular there is no memory of the initial state in the localized phase. This is an important difference with respect to the dynamics of closed many-body localized systems, which retain infinitely long lived memory of the initial state.

Under the hypothesis of Eq.~\eqref{spintempAnsatz} for ergodic phases, a natural way to fix the two 
parameters $\beta_s$ and $h$
is based on matching the expectation values of the two conserved quantities, i.e., requiring
\begin{subequations}
\label{spintemConserved}
\begin{align}
\overline{\langle \hat H \rangle}_{\mbox{\tiny stat}} &= \overline{\langle \hat H \rangle}_{\mbox{\tiny Ans}} \;, \\
\overline{\langle \hat S_z \rangle}_{\mbox{\tiny stat}} &= \overline{\langle \hat S_z \rangle}_{\mbox{\tiny Ans}}\;.
\end{align}
\end{subequations}
Here the overline represents the average over the different realizations and 
$\langle \hat O \rangle_{\mbox{\tiny stat}} = \sum_n \pnstat \bra n O \ket n$ and similarly 
$\langle O \rangle_{\mbox{\tiny Ans}} = \sum_n \pnans \bra n O \ket n$.
In Eq.~\eqref{spintemConserved}, 
the values of $s_{z,n}, \epsilon_n$ and $\pnstat$ are obtained numerically for each realization. 
Instead, the parameters $\bsst$ and $\hst$ take realization-independent values which can be solved, e.g.,  by using Newton's method.

 

\subsection{Perturbative expansion for weak interactions \label{pertpredSEC}}

A different estimation of the quasi-equilibrium parameters is based on the time-evolution of the total energy and 
magnetization. Indeed, since these two quantities are conserved by $\hat H_S$, their values merely change due to 
the spin-flip transitions induced by the 
reservoir and the microwave field. All the microscopic details are then encoded in the spin-spin correlation function,
that at large times, writes as
(see Appendix \ref{enemagnSEC})
\begin{equation}
\label{chidef}
\chi_{ij} (u,v) = \sum_n \pnstat 
\bra{n} U(u,v) \hat S_x^i 
U^\dag(u,v)  \hat S_x^j \ket{n}
\end{equation}
with $U(u,v) = e^{i (\hat H_S u + \hat S_z v)}$.
If the spin-temperature Ansatz \eqref{spintempAnsatz} holds, $\chi_{ij}(u,v)$ reduces to the calculation 
of the spin-spin correlation function at equilibrium (see Appendix \ref{perturbativeMFSEC}). 
Since, at stationarity, the total exchange of magnetization and energy  
must vanish, we obtain two conditions, which in the limit of negligible interaction strength $U$ can be written explicitly as (see Appendix \ref{relSEC})
\begin{subequations}
\label{relation01}
\begin{align}
 & \int d \omega \, f(\omega) \kappa(\omega) = 0, \label{stENE}\\
 &   \int d \omega \, \omega f(\omega) \kappa(\omega)=0.    \label{stMAGN}
\end{align}
\end{subequations}
Hereby $f(\omega)=\frac{1}{N}\sum_i\delta(\omega_e + \Delta_i -\omega)$ is the distribution of the Zeeman energies of  the electron spins, 
which we, at large $N$, chose to be uniform in $[\omega_e-\Delta\omega_e, \omega_e + \Delta\omega_e]$; $\kappa(\omega)\equiv \frac{d P_e(\omega)}{dt}$, describes the rate of change of polarization (due to radiation and reservoir) of the spins with Zeeman energy $\omega$, i.e. 
\begin{equation}
\label{relationkappa}
 \kappa(\omega) = \frac{P_0(  \omega)  -P_e(\omega) }{ 2 T_{1e}} -  \frac{T_{2e} \omega_1^2  P_e( \omega) }{T_{2e}^2 (\omega-\omw)^2+1}   \;.
\end{equation}
where $P_0(\omega)= -\tanh(\beta \omega/2)$ is the equilibrium polarization in the absence of microwaves. 
In the absence of interactions the electron polarization $P_e(\omega)$  
is fixed by $\kappa(\omega) = 0$, as different frequencies do not mix.
However, the spin-temperature Ansatz assumes the expression
$P_e(\omega) = -\tanh\bigl(\beta_s (\omega + h)/2\bigr)$ 
in the steady state: a non-zero value of $\kappa$ at specific $\omega$ is of course 
compensated by the interaction-mediated redistribution of the conserved quantities among the spins.
In Eq.~\eqref{relationkappa}, the first 
accounts for the relaxation of polarization due to the reservoir, while
the second term captures the effect of the microwave field.
A very similar form of $\kappa$ was proposed by Borghini \cite{borghini1968spin}: 
\begin{equation}
 \label{kappaborghini}
 \kappa_{\mbox{\tiny Borg}}(\omega)= \frac{P_0(  \omega)  -P_e(\omega) }{ 2 T_{1e}} -  \pi \omega_1^2  P_e( \omega)\delta(\omega - \omw)\;,  
\end{equation}
in which the Lorentzian absorption in (\ref{relationkappa}) was replaced with an infinitely sharp $\delta$-function,
assuming that only electron spins in exact resonance with $\omw$
are flipped by the microwaves. This leads, however, to a substantial underestimation of the spin-temperature \cite{Serra2012, C3CP44667K} as compared to the experimentally observed values.

Our formula in Eq.~\eqref{relationkappa} instead yields values closer to experimental observations for 
$\beta_s^{\mbox{\tiny PE}}, h^{\mbox{\tiny PE}}$.
Moreover, we can extend Eq.~\eqref{relation01} to finite $U$ and take into account perturbative corrections 
for $U\ll \Delta \omega_e$. They take the form (see Appendix \ref{relSEC}):
\begin{subequations}
\label{relation3}
\begin{align}
&   \int d \omega \, f(\omega) \left[ \kappa(\omega) + 
\frac{U^2}{2} \int d\omega' \, f(\omega') \kappa_1(\omega, \omega')\right]
  = 0,\\
&   \int d \omega \,   f(\omega) \left[ \omega \kappa(\omega) + 
\frac{U^2}{2} \int  d\omega' \,f(\omega') \kappa_2(\omega, \omega')\right]
  = 0,
\end{align}
\end{subequations}

where the second-order corrections are given by
\begin{align}
\kappa_1(\omega, \omega') &= 
 \frac{d}{d\omega} \left(\frac{\kappa(\omega) - \kappa(\omega') }{\omega - \omega'}\right), \\
\kappa_{2}(\omega, \omega') &= 
 \frac{d}{d\omega} \left(\frac{\omega \kappa(\omega) - \omega'\kappa(\omega') }{\omega - \omega'}\right).
 \end{align}

\begin{figure}[h]
\includegraphics[width=\columnwidth]{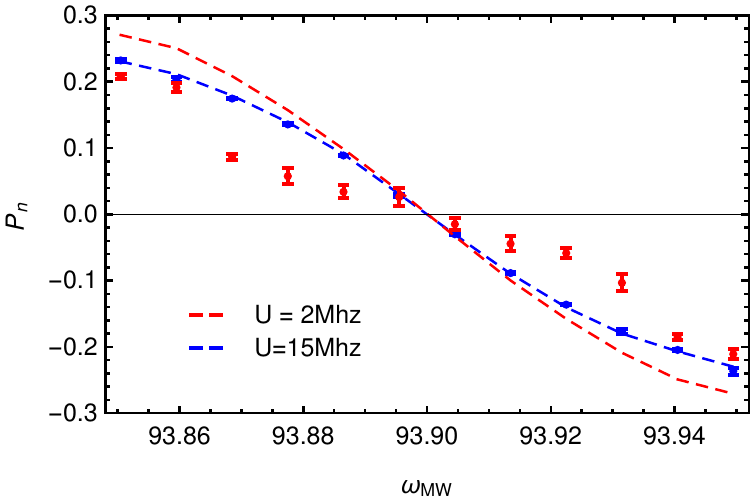}
\caption{\label{dnpprofile} 
Color Online.
\textbf{The DNP profile,} i.e. the steady state value of the nuclear polarization as a function of the microwave frequency.  In red we show the results for $U=2$ MHz, in blue the results for $U=15$ MHz. Symbols correspond to the steady state value of the nuclear polarization, while the dashed line corresponds to Eq.~(\ref{TM}) with $\beta_s$ obtained with the fitting method.}
\end{figure}

\section{Numerical results\label{secResults}}

\begin{figure*}[ht]
\includegraphics[width=0.97\columnwidth]{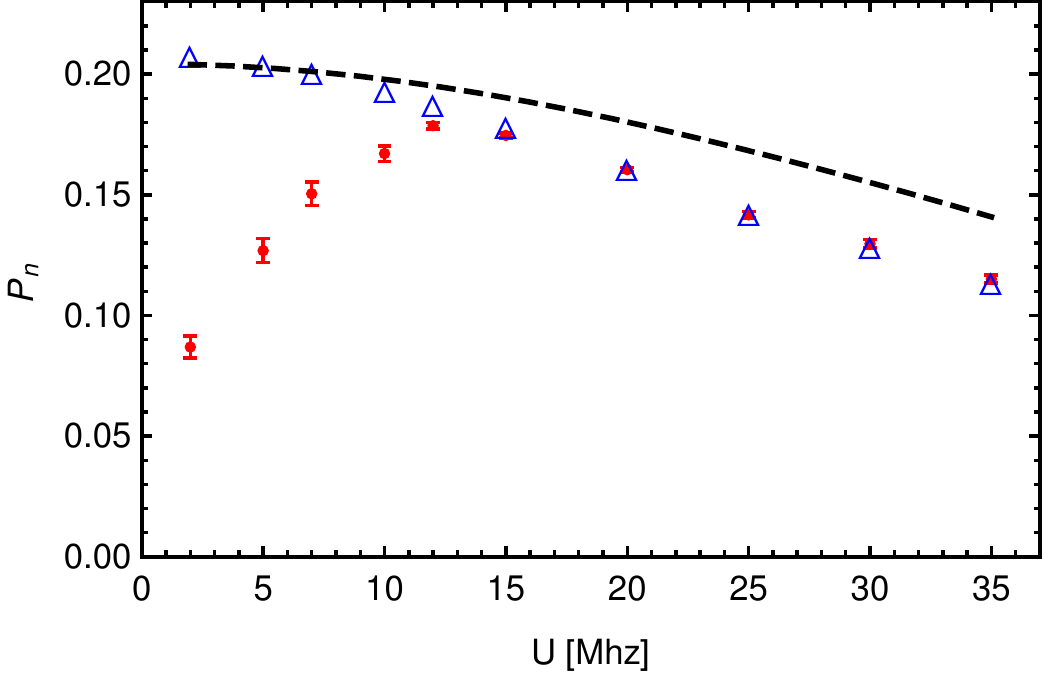}
\includegraphics[width=0.97\columnwidth]{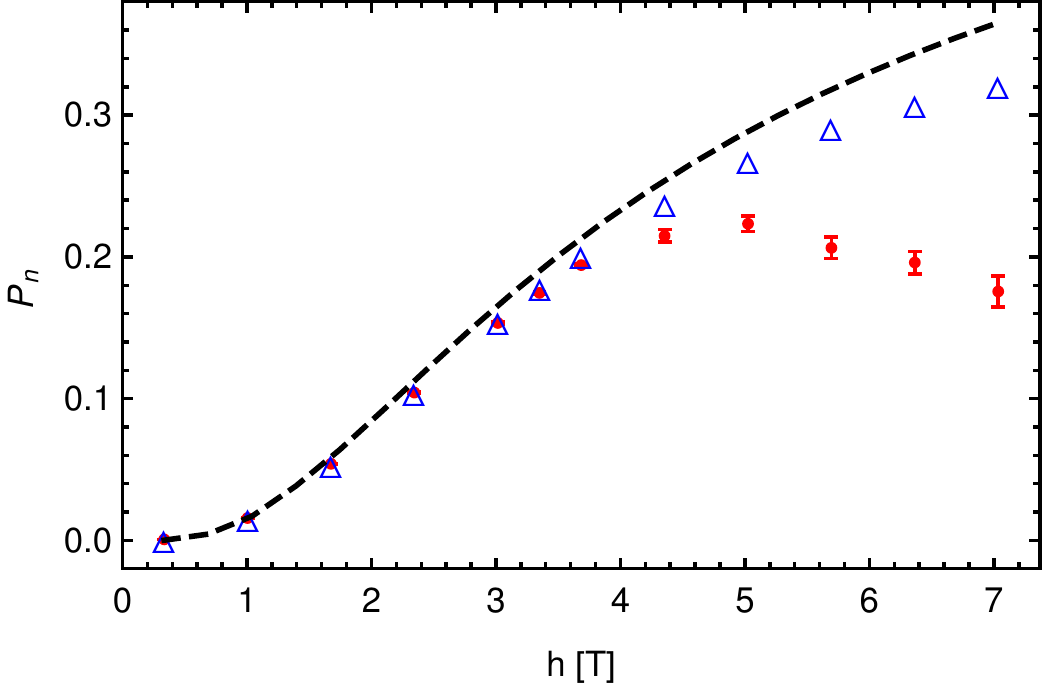}
\caption{ \textbf{Breakdown of the existence of a spin-temperature.} The steady state polarization of 
the nuclear spin is plotted as a function of the typical dipolar coupling strength $U$ at a fixed magnetic field $B=3.35 $ Tesla (Left) 
and as a function of the magnetic field $B$ at fixed $U=15$ Mhz (Right). The red disks
show the nuclear polarization in the stationary state averaged over many 
realizations (the error bar shows the standard deviation). These results are compared
with values of the nuclear polarization obtained from Eq.~(\ref{TM}).
The blue triangles correspond to a spin temperature, $\beta_s^{-1}$
 estimated by the fitting method of \eqref{spintemConserved} while the black dashed line corresponds  to the perturbative expansion discussed in Sec.~\ref{pertpredSEC}.
Both plots show the breakdown of the spin-temperature assumption once the spread of Zeeman inhomogeneities dominates over the strength of the dipolar 
interactions. 
\label{breakspin}
}
\end{figure*}

We focus on the standard conditions 
of DNP experiments using trytils for the hyperpolarization of $^{13} C$ nuclear spins.
The value of the microscopic parameters are taken from actual experiments 
and summarized in Table~\ref{tableparam}. 
\begin{table}
\begin{center}
\begin{tabular}{c|c|c|c|c}
  $T_{1e}$ &  $T_{2e}$ & $\vec{B_0}$  & $\beta$  & $U$  \\
  \hline
  $1.$ s & $10^{-6}$ s & $3.35$ Tesla &  $0.83$ K $^{-1}$ & $2.0 \div 45.0$ MHz \\
\end{tabular}\\
\vspace{0.2cm} 
\begin{tabular}{c|c|c|c}
  $\omega_e$  & $\Delta\omega_e$ & $\omega_1$ & $\omega_n$  \\
  \hline
  $93.9\ 2\pi$GHz & $54\ 2\pi$GHz  & $0.25 \times 10^{-4}\ 2\pi$GHz  & $20\ 2\pi$MHz  \\
\end{tabular}
\caption{\label{tableparam} Summary of the parameters used in our calculations.}
\end{center}
\end{table}

The aim of this section is twofold: on the one hand,
we test the range of validity of the spin-temperature assumption for the stationary state; 
on the other hand, we quantify the finite-size correction affecting our numerical results for $N = 12$ (averaged over at least
$100$ configurations). 
To achieve this, we focus on the stationary nuclear spin polarization.
The nuclear spin is weakly coupled with the electron spins and acts
simply as a thermometer; within the spin-temperature Ansatz, its stationary polarization is therefore expected to take the form of Eq.~(\ref{TM}).
We can then compare three different estimations for $P_n$:
\begin{itemize}
 \item the value predicted by Eq.~\eqref{TM}, with $\beta_s = \bsst$ as obtained from the fitting method
  explained in Sec.~\ref{spintempSEC};
 \item the value predicted by Eq.~\eqref{TM}, with $\beta_s = \beta_s^{\mbox{\tiny PE}}$, obtained from the perturbative
 expansion in the  thermodynamic limit $N\to \infty$ of the fully-connected model, as explained in Sec.~\ref{pertpredSEC};
 \item the exact value obtained upon averaging over several realizations the stationary nuclear spin polarization obtained from the numerical procedure
 over several realizations
 \begin{equation}
 \label{pnstat}
P_n = \overline{\sum_m p_m^{\mbox{\tiny stat}} \bra{m} I_z \ket{m}} \;.
\end{equation}
\end{itemize}
In Fig.~\ref{breakspin} the dashed lines are the prediction
for the nuclear polarization with the analytical estimates for the spin-temperature. 
For small interaction, the perturbative results of Eqs.~\eqref{relation3} are 
in good agreement with the numerical data.
Upon increasing $U$, the lowest order result Eq.~(\ref{relation3}) cannot be expected to be accurate anymore, 
but it still correctly describes the decrease in polarization, and thus captures the important, 
but hitherto unexplained, effect of radical concentration seen in the experiments \cite{ardenkjaer2003increase, ColomboSerra2014}: 
as the radical concentration, and thus $U$, is increased, 
the nuclear polarization decreases. 


\begin{figure*}[ht]
\includegraphics[width=0.8\textwidth]{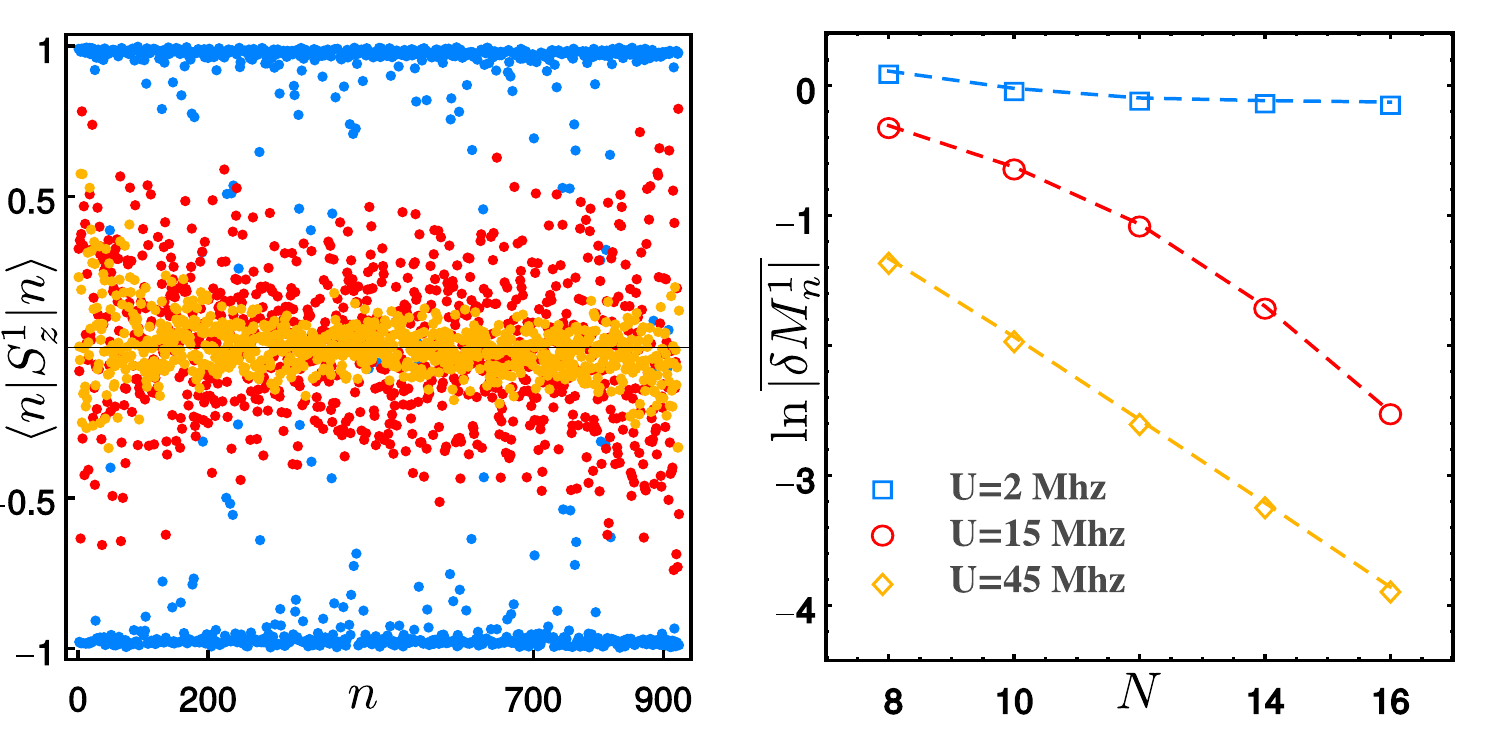}
\caption{\label{ETHsketch}
Color online. Numerical investigation of the ETH hypothesis for a system of $N$ electron spins. (a) 
Average polarization of the selected spin $\hat S_z^1$ in the eigenstates $\ket{n}$ of $N=12$ electron spins with vanishing total magnetization. The $\ket{n}$ are ordered according to increasing energy. Data are shown for three different values of the dipolar strength:
$U = 2$ MHz (blue), $15$ MHz (red), $45$ MHz (blue); $\Delta \omega_e=54$ MHz.
(b) Difference
in local magnetization between consecutive eigenstates, $\delta M^1_n = \bra{n+1} \hat S_z^1 \ket{n+1}-\bra{n} \hat S_z^1 \ket{n}$, 
averaged over disorder realizations and globally unpolarized eigenstates, for different system sizes $N = 8,\ldots, 16$. An exponential decay of $\overline{\delta M^1_n}$ with $N$ indicates a thermal phase obeying ETH, while a saturation signals many-body localization.}
\end{figure*}

On the left of Fig. \ref{breakspin}, we show the behavior of the nuclear hyperpolarization as a function of the
dipolar coupling $U$, at fixed disorder strength. We observe two regimes: in the strongly-interacting regime,
the concept of a spin-temperature perfectly applies to the stationary states and the polarization of the nuclear spin 
collapses with the prediction given by \eqref{TM} and $\beta_s = \bsst$. For $U \lesssim 10 $ MHz, ($U/\Delta\omega_e \lesssim 0.2$)
the observed polarization is much smaller than the value expected by postulating a spin-temperature state of the electrons. 
This shows that at least for our finite $N$ simulations, the thermal regime breaks down. We expect that this behavior remains
true in the thermodynamic limit of systems with a finite connectivity among electron spins.
A similar behavior is observed in the right panel of Fig.~\ref{breakspin}, where the magnetic field is varied at constant interaction strength $U$. We recall that
the disorder strength, i.e., the spread of inhomogeneous contributions to the Zeeman energies, $\Delta\omega_e$, 
increases proportionally with the external magnetic field $|\vec B|$. 
Thus, at small fields the dominating  interactions establish a  spin-temperature, whereas at large fields, the internal thermalization of the electron system breaks down. 
We note that both methods to compute the spin-temperature, 
the fitting procedure for $N=12$ systems and the perturbative analytical calculation for 
a mean field ($N=\infty$) system, yield compatible results. Their difference seems to stem mostly 
from the error due to the restriction of the perturbative calculation to quadratic order in $U$,
rather than due to the effects of comparing $N=12$ with the mean field limit $N=\infty$.

In Fig.~\ref{dnpprofile},  we show an important characteristics of a DNP experiment, known as the DNP profile: the hyperpolarization as a function of 
the irradiated microwave frequency $\omw$. We compare the exact value of $P_n$ from \eqref{pnstat} with
the prediction from the fitting method and plugging $\beta^{\rm FM}$ into \eqref{TM}.
For $U=15$ MHz ($U/\Delta\omega_e \simeq 0.3$), the two results are consistent, while
for $U=2$ MHz ($U/\Delta\omega_e\simeq 0.04$), we generally observe a small hyperpolarization, with the exception of the 
a window of width $O(\omega_n)$ around $\omega_{\rm MW}=\omega_e\pm \Delta\omega_e$, 
where
the polarization is induced by the so-called solid effect \cite{Karabanov2012}. The latter consists in the following: Even in the absence of dipolar interactions among the electrons,
the presence of hyperfine interactions allows the microwaves to excite an prohibited transition, where the nuclear spin is flipped together
with an electron with a transition frequency $\omega_i = \omw \pm \omega_n$. This effect is more prominent at the boundaries
of the microwave spectrum of interest, $\omw \in [\omega_e- \Delta\omega_e,\omega_e+ \Delta\omega_e]$. Indeed, in the bulk of the spectrum simultaneous flips of an electron and a nuclear spin have similar probability, and therefore their effects tend to cancel. 


\section{Discussion} \label{secdiscussion} 
As discussed in the previous section, 
the data presented in Fig.~\ref{breakspin} indicate a marked change of behavior for weak dipolar couplings (i.e., for low radical concentration) and/or in large magnetic fields. 
Since the hyperfine interaction between the nuclei and the electrons
is left unchanged in both these cases, this phenomenology actually reflects a change 
in the electron system. Indeed, in order for the electron spins
to act as a bath for the nucleus, they have to be in a "thermal phase". 
Upon decreasing the value of $U$, the disorder in the inhomogeneous Zeeman energies, $\Delta \omega_e$, becomes dominant: 
the eigenstates fail to be ergodic and enter a many-body localized phase. 
Indications of this transition are shown in Fig.~\ref{ETHsketch} where we study
the expectation value of the polarization of a selected electron spin, say $S^1$, on all eigenstates
within the sector of vanishing total electron polarization: $\bra{n} \hat S_z^1 \ket{n}$.
The left of Fig.~\ref{ETHsketch} visualizes the qualitatively different behaviors for weak and strong interactions, respectively, in a single realization, at fixed magnetic field $B=3.35$ Tesla:
At weak interactions $U \simeq 2$ MHz ($U/\Delta \omega_e \simeq 0.04$), the expectation value $\bra{n} \hat S_z^1 \ket{n}$ fluctuates between the fully polarized extremes $\pm 1/2$. 
In contrast, when the interactions dominate, $U \simeq 15, 45$ MHz,  $(U/\Delta \omega_e \simeq 0.3, 0.8)$,
the values of $\bra{n} \hat S_z^1 \ket{n}$ instead concentrate close to the equilibrium value $0$. 
This difference becomes sharper and sharper with increasing system size, as analyzed by the average variations of the local magnetization as a function of $N$, c.f. the right panel of Fig.~\ref{ETHsketch}. The extrapolation of such data to the thermodynamic limit allows one to locate the many-body localization transition.
The data confirm that at strong interactions $U$ there is a thermal phase obeying ETH. This is exemplified by the expectation values of the local observable $S^1_z$ which coincide for all eigenstates and agree with the thermodynamic average. 
In contrast, for small $U$ and the accessible system sizes, 
we observe the fingerprints of many-body localization, with strong fluctuations of local observables between different eigenstates.
In this case, thermodynamic expectation values can only be recovered by averaging over many eigenstates.

\begin{figure*}[ht]
\includegraphics[width=0.97\columnwidth]{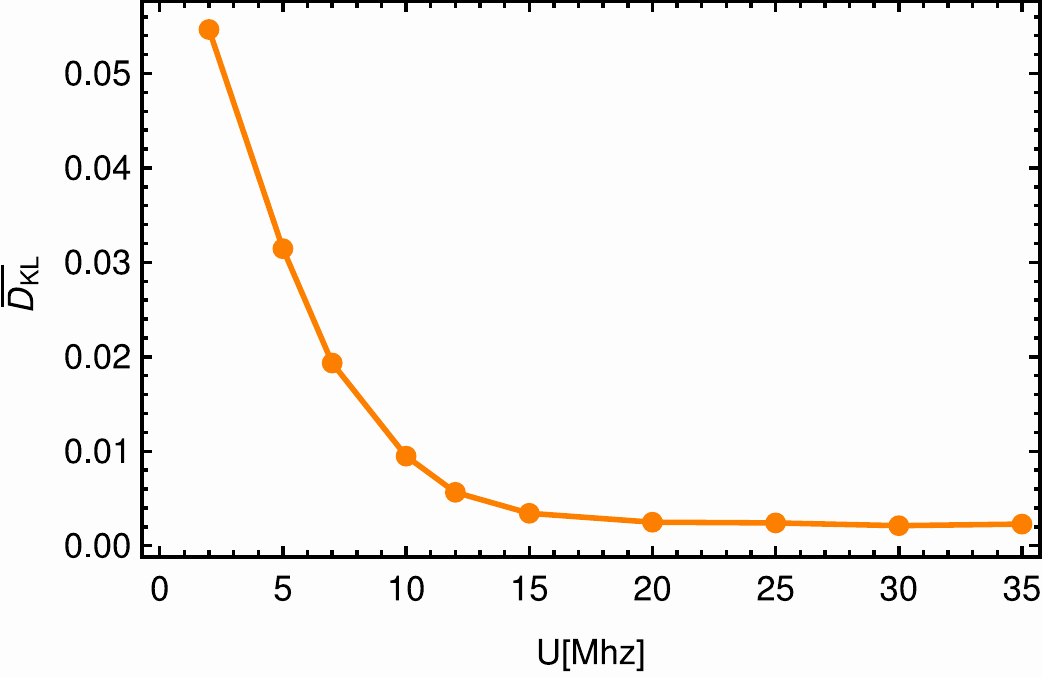}
\includegraphics[width=0.97\columnwidth]{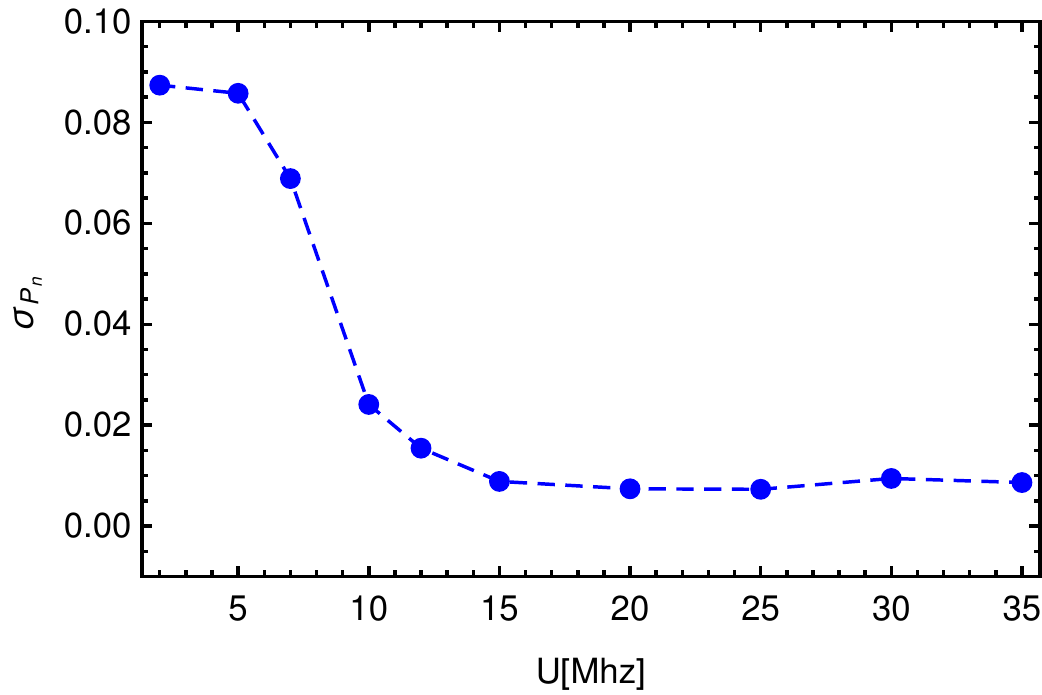}
\caption{(a) 
The Kullback-Leibler divergence, comparing the distribution in the stationary state $\pnstat$ with the spin-temperature
Ansatz $\pnans$ of \eqref{spintempAnsatz}.
(b) The sample-to-sample fluctuations of the nuclear polarization 
as a function of $U$. The localized phase is characterized by 
vastly enhanced fluctuations of the polarization of the probing nuclear spin, 
demonstrating the absence of a homogeneous spin temperature in the electronic system. \label{mblfig}
}
\end{figure*}

Internal thermalization among the electron spins explains the emergence of an effective spin-temperature in DNP experiments, as long as the thermalization is much faster than the driving and bath relaxation processes. In the ergodic phase this is essentially always the case, given that the latter given that the latter are orders of magnitude slower than the internal spin dynamics. 
Indeed, the thermal eigenstates cannot encode any memory about local perturbations such as the frequency-selective
spin-flips driven by the microwave radiation. In contrast, in the localized phase, 
there exists an extensive set of local conserved quantities in the isolated system. This implies that  two parameters $\beta_s$ and $h$
cannot contain sufficient information to describe local observables in all eigenstates of given total energy and polarization. 
As a consequence, the steady state and its properties will be more complex, and depend on details of how the radiation and the bath couple to individual spins, and how those are coupled among each other. 
Our numerical results show that the steady state concentrates on eigenstates 
with a vanishing polarization of resonant electron spins and a strong 
polarization of the non-resonant ones.  

The dependence of the nuclear spin polarization (via the spin temperature) on interaction strength and magnetic field, as shown in Fig.~\ref{breakspin}, can be understood, at least at a qualitative level.
Indeed, when the ratio between interactions and magnetic field, $U/B$, is increased,
two competing effects are enhanced simultaneously: 
\begin{itemize}
 \item The tendency towards thermalization increases, which eventually leads to ETH  and the appearance of
 the spin-temperature;
 \item The microwave irradiation is effective on a larger number of spins and thus acts less selectively. Indeed, all spins with Zeeman gap satisfying $|\omega_e+\Delta_i - \omega_{MW}| = T_{2e}+ O(\delta E(U))$ will absorb the microwave irradiation efficiently. Hereby, 
$\delta E(U) \simeq \min [U^2/\Delta \omega_e, U]$ is the interaction-induced width of the local spectral functions of typical spins. The associated broadening of the absorption line implies a broader range of spins  with suppressed polarization in the steady state, and thus an increase in the resulting spin temperature
\end{itemize}
In practice, while a sufficiently large $U$ is needed to ensure thermalization among the electron spins, too large a value
broadens the absorption line, which ultimately results in a stationary state with a higher spin temperature.
Therefore, we reach the conclusion that the inverse spin temperature, and thus the achieved hyperpolarization level of nuclear spins,
will reach a maximal value when $U/B$ is tuned to the proximity of the many-body localization transition: There,
thermalization still occurs, but the microwave irradiation couples to spins in a maximally narrow frequency range, enabling a low spin temperature to emerge.

Additional insights of the validity or failure of the spin-temperature Ansatz for different values of $U$
are provided in Fig.~\ref{mblfig}.
It is possible to give a quantitative estimation of the validity of the Ansatz in \eqref{spintempAnsatz} 
using a standard statistical indicator, known as Kullback-Leibler divergence.
It quantifies the amount of information loss when $\pnans$ is used to approximate $\pnstat$
and is defined as 
\begin{equation}
 D_{KL}  = \sum_n \ln \left(\frac{\pnstat}{\pnans}\right) \pnstat 
\end{equation}
The average of $D_{KL}$ over disorder realizations is shown in Fig.~\ref{mblfig} left, as a function of the dipolar coupling. 
For small values of $U$, the large value of $D_{KL}$ reflects the fact that
the spin-temperature Ansatz cannot 
reproduce accurately the local behavior of the spin-polarizations. At large $U$, where the spin-temperature picture applies,
the value of $D_{KL}$ becomes much smaller. An interesting question is whether a small but finite difference 
always remains in the thermodynamic limit, i.e. 
if with an appropriate observable the stationary state could be distingushed from thermal ensemble.

A fingerprint of the localization transition is seen in the sample-to-sample fluctuations of 
the stationary nuclear polarization shown in Fig.~\ref{mblfig} right.
In the thermal phase, fluctuations originate only from the finite-size effects on the total energy and magnetization 
in the stationary state and are therefore largely suppressed. In the localized phase instead, local processes
control the nuclear polarization and the final polarization depends on the presence or absence of few-body resonances. 

\section{Conclusion}
We presented a simple model for an improved description of DNP, which accounts for the crucial role played by dipolar interactions among radical spins.
For sufficiently strong dipolar coupling, the electron spin system
thermalizes internally and acts as an effective thermal bath 
for the nuclear spins and cools them down to the spin-temperature established among the electron spins.
For this regime, we analytically estimated the achieved nuclear spin polarization in a perturbative expansion in the dipolar coupling, which we nevertheless assume strong enough to ensure a thermal phase. 
 
In that phase the maximal hyperpolarization in DNP is obtained by minimizing the effective spin temperature. Our analysis shows that this is achieved by reducing the interaction strength to maximal possible extent, that is
upon approaching the localization transition from the ergodic phase. This can be achieved by either reducing the radical concentration or increasing the magnetic field strength. The localized phase of the electron spins instead yields a significantly lower degree of nuclear spin polarization.
Here a word of caution qualifying the meaning of many-body localization in real dipolar systems is in order. A genuine localization transition in the isolated system at finite energy density is expected only if the interactions are sufficiently short range. For dipolar interactions, already Anderson's work~\cite{Anderson1958} suggested delocalization, even though on exponentially long time scales, as $U/\Delta \omega_e$ becomes small. This is because dipolar couplings decay as a marginal power law, which always allow spins to find resonant partners at large distance. 
Even in $d=2$ dimensions it has been argued that there are channels for delocalization~\cite{burin2006energy, yao2014many}, 
with however, even longer time scales. When invoking a localization transition, we in fact allude to a strong crossover to very
long thermalization time scales which grow exponentially with $\Delta \omega_e/U$. The latter makes the spin temperature 
Ansatz break down rather quickly, too.
Similarly, the mean field model is not expected to have a genuine localization transition as $N\to \infty$, which further justifies a posteriori our expansion around the non-interacting limit. Nevertheless, a thorough 
understanding of the thermalization processes and the involved time scales in the limit $N\to \infty$ require further analysis.

It would be interesting to test these predictions in standard experimental DNP set-ups, both to validate
our conclusions and to investigate the manifestation of the many-body localization transition
in out-of-equilibrium stationary states. 

It will be important to understand the crossover to a regime where the driving is eventually faster than the dephasing and/or the internal equilibration times, so that hole-burning phenomena and saturation of the driving efficiency become important. 
Furthermore, one should understand more thoroughly the localization transition as a function of the effective connectivity of the electron spins, which presumably varies with their dilution and the associated positional randomness in the radical spins. 
The latter presumably varies with the dilution of radicals and their positional randomness.
We plan to address these issues in a future publication.

\begin{acknowledgments}
This work is supported by ``Investissements d'Avenir'' LabEx PALM 
(ANR-10-LABX-0039-PALM).
We thank L. Mazza and Xiangyu Cao for interesting discussions,
S. Colombo Serra for the support with experimental data and
C. Zankoc for collaboration in the early stage of this work.
\end{acknowledgments}

\bibliography{dnp}

\clearpage
\newpage

\setcounter{page}{1}
\clearpage 

\onecolumngrid

\appendix

\section{Borghini model with interactions \label{enemagnSEC}}
In this appendix, we provide a general method to determine the two parameters $\beta_s, h$ conjugate to the two conserved quantities of the Hamiltonian $\hat H$ in \eqref{hamiltonian}. 
In the dynamics described by the master equation \eqref{mastereq1},
the system exchanges simultaneously energy and magnetization every time a
spin is flipped by the reservoir or the microwaves. 
We can compute explicitly the joint probability distribution 
$P(\Omega, S)$ of the energy and magnetization variation within a time interval $[t, t+ \delta t]$: 
\begin{equation}
\label{PWSdef}
 P(\Omega,S) = \sum_{n n'} \delta[\Omega - (\epsilon_n - \epsilon_{n'})] \delta[S - (s_z^n - s_z^{n'})]  P_{n' \to n}^{\delta t} p_{n'}(t)\;, 
\end{equation}
where $P_{n'\to n}^{\delta t}$ is the probability of passing from $n' \to n$ in a time $\delta t$. In the limit $\delta t \to 0$, we have therefore
for $n \neq n'$, $P_{n'\to n}^{\delta t} = W_{n' \to n} \delta t$ and $P_{n \to n}^{\delta t} = 1 - \sum_{n' \neq n} P_{n \to n'} = 1 - \sum_{n'\neq n} W_{n\to n'} \delta t$. Replacing in \eqref{PWSdef}, 
we arrive at $P(\Omega,S,t) = P_{n\to n}\delta(\Omega)\delta(S) + \delta t\, p(\Omega,S,t)$ with

\begin{equation}
\label{pWSdef}
 p(\Omega,S,t) = \sum_{\substack{n,n' \\ n\neq n'}} \delta[\Omega - (\epsilon_n - \epsilon_{n'})] \delta[S - (s_z^n - s_z^{n'})]  W_{n' \to n} \; p_{n'}(t) \;, 
\end{equation}
while the term $P_{n\to n}$ does not contribute to the average of $\Omega$ and $S$. 
In the large time limit $p_n(t) \to \pnstat$; moreover,
using the rates in (\ref{matrixtransitionBATH},~\ref{matrixtransitionMW})  and the integral representation of the $\delta$-function, we have
\begin{equation}
\label{bathterm}
\sum_{n\neq n'} \delta[\Omega - (\epsilon_n - \epsilon_{n'})] \delta[S - (s_z^n - s_z^{n'})] W^{\text{bath}}_{n, n'} \pnstatp = 
\sum_{j=1}^N \int \frac{du}{2\pi} \frac{dv}{2\pi} e^{i u \Omega + i v S} \frac{4 h_{\beta}(-\Omega)}{T_{1e}} \chi_{jj}(u,v) \;, 
\end{equation}
where $\chi_{jj}(u,v)$ is defined in \eqref{chidef} and 
we used rotational symmetry around the $z$-axis. 
Analogously for the microwave rate, we obtain
\begin{equation}
\label{mwterm}
\sum_{n\neq n'} \delta[\Omega - (\epsilon_n - \epsilon_{n'})] \delta[S - (s_z^n - s_z^{n'})] W^{\text{MW}}_{n, n'} \pnstatp= 
\sum_{ij}\int \frac{du}{2\pi} \frac{dv}{2\pi} e^{i u \Omega + i v S} \frac{4 \omega_1^2 T_{2e}}{1 + T_{2e}^2 (\Omega - \omw S)^2} \chi_{ij}(u,v) \;.
\end{equation}
Note that in order to write Eqs.~(\ref{bathterm}, \ref{mwterm}), we used explicitly that $\ket{n}$ is a simultaneous eigenstate
of $\hat H$ and $\hat S_z$: so this derivation only holds for the conserved quantities of the model. 

Let us introduce the Fourier transform of the correlation function
\begin{equation}
\label{FourierAB}
\chi_{ij}(\Omega, S) = \int \frac{du}{2\pi}\frac{dv}{2\pi} e^{i (\Omega u + S v)} \chi_{ij}(u,v) \;, 
\end{equation}
with which, we can rewrite equation \eqref{pWSdef} as
\begin{equation}
\label{pWSfour}
  p_{stat}(\Omega, S) = \frac{4 h_{\beta}(-\Omega)}{T_{1e}} \sum_{j} \chi_{jj} (\Omega, S) + \frac{4 \omega_1^2 T_{2e}}{1 + T_{2e}^2 (\Omega - \omw S)^2}
 \sum_{ij} \chi_{ij} (\Omega, S)\;.
\end{equation}
Imposing that the energy and magnetization flows vanish in the stationary state we find two equations
\begin{equation}
\label{floweq}
 \int dS d\Omega \, p(\Omega, S) \Omega = 0\;,\qquad \int dS d\Omega \, p(\Omega, S) S = 0 \;.
\end{equation}

If we assume that the stationary state is effectively thermal, in agreement
with \eqref{spintempAnsatz}, i.e., $\pnstat = e^{-\beta_s (\epsilon_n + h s_{z,n})}/Z$, Eqs.~\eqref{floweq} suffice to determine 
the two parameters $\beta_s, h$. The essential ingredient is the correlation function $\chi_{ij}(u,v)$, which is derived in the next appendix.

\section{Perturbative calculation of the correlation function in the mean-field model \label{perturbativeMFSEC}}
We now compute the correlation function $\chi_{ij}(u,v)$ to second order in the dipolar interaction strength $U$. It is analyzed in the mean-field model defined in \eqref{hamiltonian}, 
in the thermodynamic limit $N\to \infty$. 
We keep the inhomogeneities $\Delta_i$ finite and arbitrary.
As the nuclear spin only slightly perturbs the electron spin Hamiltonian, we can neglect it in the estimation of $\beta_s$ and  $h$ and focus on the subsystem of interacting electron spins:
\begin{equation}
 \label{HamiltonianElectron}
  \hat H_e(h) =  \sum_{i =1}^N \left( \omega_e + h +\Delta_i \right) \hat S_z^i 
+ \sum_{i<j} \dip_{ij} (\hat{S}^i_+ \hat{S}^j_- +\hat{S}^i_- \hat{S}^j_+)=\hat{H}_0+\hat{V}\;,
\end{equation}
where $\hat{H}_0$ is the non-interacting Hamiltonian, and $\hat{V}$ describes the dipolar couplings $\dip_{ij}$ between electron spins, which are Gaussian random variables with covariance matrix 
\begin{equation}
\label{covariance}
\overline{\dip_{ij}\dip_{kl}}=\frac{U^2}{N}\delta_{ik}\delta_{jl}\;.
\end{equation}
Note that, at order $U^2$, the correlation functions for different spins will be decoupled 
after averaging because of (\ref{covariance}), i.e.,  
$\chi_{i j}(\Omega, S) = \delta_{ij}\chi_{i i}(\Omega,S) + o(U^2)$. Thus, we can just restrict ourselves to the calculation of 
the dynamical correlation function $\chi_{j j} (u,v)$ of a single-spin $j$. It can be simplified as
\begin{equation}
\label{xxpm}
\chi_{jj}(u,v) = \frac{e^{i (v - hu)} \Gamma_+^j(u, h) + e^{- i (v - hu)} \Gamma_-^j(-u, h)}{4}
\end{equation}
by introducing the $S^j_+ S^j_-$ correlators
\begin{equation}
\label{Cplus}
\Gamma_{\pm}^{j} (u,h) = Z^{-1}\Tr [e^{-(\beta_s - i u)\hat{H}_e(h)}  \hat{S}_\pm^j e^{-i u \hat{H}_e(h)} \hat{S}_\mp^j] \;.
\end{equation}
For simplicity, in the following we will keep tacit the dependence on $h$,
since it simply amounts to changing $\omega_e \to \omega_e + h$ in $\hat H_e(0)$. 
In order to take advantage of the path integral formalism, we perform a Wick rotation and define $C_{\pm}^i(\tau) = \Gamma_{\pm}^i(- i \tau)$.
From the cyclicity of the trace, we deduce $C_{-}^i (\tau) = C_{+}^i(\beta_s - \tau)$, and 
we can thus restrict ourselves to calculating $C_{+}^i(\tau)$. 
We introduce
the evolution operator in the interaction picture as
$\hat U_\tau = e^{\tau \hat H_0} e^{-\tau \hat H}$, so that
\begin{equation}
 \frac{d \hat U_\tau}{d\tau} = \hat H_0 \hat U_\tau - e^{\tau \hat H_0} \hat H_e e^{-\tau \hat H_0} \hat U_\tau
  = -\hat V_{I}(\tau) \hat U_\tau \;,
\end{equation}
where $\hat V_{I}(\tau) = e^{\tau \hat H_0} \hat V e^{-\tau \hat H_0} $. Integrating this last equation over $[0,\tau]$ 
and re-injecting the resulting equation to itself, we get the second order expansion 
\begin{equation}
 e^{-\tau\hat H} =  e^{-\tau\hat H_0}  - \int_0^\tau dt' e^{(t' - \tau)\hat H_0}\hat V e^{-t' \hat H_0} + 
 \int_0^\tau dt' \int_0^{t'} dt'' e^{(t' - \tau) \hat H_0} \hat V e^{(t'' - t') \hat H_0} \hat V e^{-t'' \hat H_0} \;.
\end{equation}
In order to keep track of all the terms in the expansion in $U$, we write $\hat V \to \epsilon \hat V$. We can expand
the reduced partition function and the correlation function as
\begin{align}
\frac{Z}{Z_0} &= ( 1 + \epsilon z^{(1)} + \epsilon^2 z^{(2)} ) \;,
\\
\frac{C_{+}^{i} (\tau) Z}{Z_0} &= c_{+}^{i,(0)}(\tau) + \epsilon c_+^{i,(1)}(\tau) + \epsilon^2 c_+^{i,(2)}(\tau) \;,
\end{align}
where $Z_0$ is the non-interacting ($\dip_{ij} = 0$) partition function. Then, we have the expansion
\begin{equation}
\label{expansionsecondorder}
 \overline{C_{+}^{i}} = c_{+}^{i,(0)}(\tau)(1 + 
 \overline{(z^{(1)})^2} - \overline{(z^{(2)})}) - \overline{z^{(1)} c_+^{i,(1)}(\tau)} + \overline{c_+^{i,(2)}(\tau)} \;,
\end{equation}
where the overline represents the average taken over the distribution of the dipolar couplings $\dip_{ij}$, with fixed values of the $\Delta_i$'s.
Note that $\overline{\hat V} = 0$ so the first non-vanishing correction is quadratic. Moreover $z^{(1)}$ vanishes identically since $\bra{n_0} V \ket{n_0} = 0$ for 
any eigenstate $\ket{n_0}$ of $\hat H_0$.
After some algebra, the calculation up to second order of each term leads to the expression
\begin{equation}
\label{finalCp}
 \overline{C_{+}^{i} (\tau) } = c_{+}^{i,(0)}(\tau) + \frac{U^2}{N}\sum_{l \neq i} \frac{d}{d\Delta_i}
 \left[\frac{c_{+}^{i,(0)}(\tau)-c_{+}^{l,(0)}(\tau)}{\Delta_i - \Delta_l}\right]\;,
 \end{equation}
where 
\begin{align}
\label{cp0}
c_{+}^{i,(0)}(\tau) &= \frac{\Tr[e^{-(\beta_s - \tau) (\omega_e + \Delta_i) S_z} S^i_+ e^{- \tau (\omega_e + \Delta_i) S_z}
 S^i_-]}{\Tr[e^{-\beta_s (\omega_e  + \Delta_i) S_z} ]} = \frac{e^{\tau (\omega_e+\Delta_i)}}{1 + e^{\beta_s (\omega_e +\Delta_i)}}\;.
\end{align}
A simpler way to derive Eq.~\eqref{finalCp} is to note, by simple power counting, that at  order $U^2$,
the perturbation in $ \overline{C_{+}^{i} (\tau) }$ 
must consist in additive contributions from all spins $l \neq i$. The term inside
the sum in \eqref{finalCp} can then be obtained 
by solving exactly the case $N = 2$ for an arbitrary value of the coupling $\dip_{12}$ in \eqref{HamiltonianElectron}
and expanding up to the order $\dip_{12}^2$. 

Note that the expression \eqref{cp0} allows us also to compute the local polarizations:
\begin{equation}
 P_i = 2\Tr[ S_z^i \rho ] = 
 \Tr[ S_+^i S_-^i \rho ] -  \Tr[ S_-^i S_+^i \rho ] = \overline{C_{+}^{i} (0) } - 
\overline{C_{-}^{i} (0) } = 
P_i^0 + \frac{U^2}{N}\sum_{l \neq i} \frac{d}{d\Delta_i}
 \left[\frac{P_i^0-P_l^0}{\Delta_i - \Delta_l}\right]\;,
 \end{equation}
where $P_i^0 = - \tanh\left(\frac{\beta_s (\omega_e + \Delta_i)}{2}\right)$ are the non-interacting polarizations
for $U=0$. 

Performing back the Wick rotation $\tau = i u$ and replacing $\omega_e \to \omega_e + h$ in \eqref{finalCp} and \eqref{cp0},
we obtain $\Gamma_{\pm}^i(u,h)$ 

\begin{align}
\Gamma_{+}^i(u,h)=\frac{e^{i u \left(\omega _e+h+ \Delta _i\right)}}{1+e^{\beta_s \left(\omega _e+h+\Delta _i\right)}} &+
\frac{U^2}{N} \sum_{l \neq i} \frac{d}{d\Delta_i}
 \left[\frac{ \frac{e^{i u \left(\omega _e+h+ \Delta _i\right)}}{1+e^{\beta_s \left(\omega _e+h+\Delta _i\right)}} - 
\frac{e^{i u \left(\omega _e+h+ \Delta _l\right)}}{1+e^{\beta_s \left(\omega _e+h+\Delta _l\right)}}}{\Delta_i - \Delta_l}\right] \;,
\\
\Gamma_{-}^i(u,h)=\frac{e^{(\beta_s - i u) \left(\omega _e+h+ \Delta _i\right)}}{1+e^{\beta_s \left(\omega _e+h+\Delta _i\right)}} &+ 
\frac{U^2}{N} \sum_{l \neq i} \frac{d}{d\Delta_i}
 \left[\frac{ \frac{e^{(\beta_s - i u) \left(\omega _e+h+ \Delta _i\right)}}{1+e^{\beta_s \left(\omega _e+h+\Delta _i\right)}} - 
\frac{e^{(\beta_s - i u) \left(\omega _e+h+ \Delta _l\right)}}{1+e^{\beta_s \left(\omega _e+h+\Delta _l\right)}}}{\Delta_i - \Delta_l}\right] \;.
\end{align}

\section{Equations for $\beta_s$ and $h$ in the mean-field model \label{relSEC}}

Using \eqref{cp0} and \eqref{FourierAB}, the correlation $ \chi_{jj}^{U=0}(\Omega, S)$ in the non-interacting case is:
\begin{equation}
 \chi_{jj}^{U=0}(\Omega, S) =
\frac{\delta (S-1) h_{\beta_s} (\omega_e + h + \Delta_j) \delta (\Delta_j -\Omega +\omega_e)}{4}+
\frac{\delta (S+1) h_{\beta_s} (-\omega_e - h - \Delta_j)\delta (\Delta_j +\Omega + \omega_e)}{4}\;.
 \end{equation}
Using \eqref{pWSfour} and \eqref{floweq}, the integration over $S,\Omega$ leads to the equations which fix $h$ and $\beta_s$:
\begin{align}
 & \sum_j \kappa(\omega_e + \Delta_j)=0 \;,
 \\
 & \sum_j (\omega_e + \Delta_j) \kappa(\omega_e + \Delta_j) =0 \;,
 \end{align}
where $\kappa(\omega)$ is defined in \eqref{relationkappa}. Eqs.~\eqref{relation01} 
are recovered in the large $N$ limit, where the sums over $j$ 
can be converted into integrals over the distribution of inhomogeneities, $f(\Delta)$.
Then, since \eqref{finalCp} is a linear combination of the functions the $c_+^{i,(0)}(\tau)$, to order $U^2$ the equations simply become
\begin{align}
 & \sum_j \left[\kappa(\omega_e + \Delta_j) + \frac{U^2}{N} \sum_{l\neq j} 
 \frac{d}{d\Delta_j} \left(\frac{\kappa(\omega_e + \Delta_j) - \kappa(\omega_e + \Delta_l) }{\Delta_j - \Delta_l}\right)\right]=0 \;,
\label{eqmagnU2}
 \\
 & \sum_j \left[ (\omega_e + \Delta_j) \kappa(\omega_e + \Delta_j) + \frac{U^2}{N} \sum_{l\neq j} 
 \frac{d}{d\Delta_j} \left(\frac{(\omega_e + \Delta_j) \kappa(\omega_e + \Delta_j) - (\omega_e + \Delta_l) \kappa(\omega_e + \Delta_l) }{\Delta_j - \Delta_l}\right)\right]=0 \;.
 \label{eqeneU2}
 \end{align}


\end{document}